# Single laser pulse induced magnetization switching in in-plane magnetized GdCo alloys

Jun-Xiao Lin,[1] Michel Hehn,[1,2] Thomas Hauet,[1] Yi Peng,[1] Junta Igarashi,[1] Yann Le Guen,[1] Quentin Remy,[3] Jon Gorchon,[1] Gregory Malinowski,[1] Stéphane Mangin,*[1,2] and Julius Hohlfeld[1]

[1]*Université de Lorraine, CNRS, Institut Jean Lamour, F-54000 Nancy, France*
[2]*Center for Science and Innovation in Spintronics, Tohoku University, Sendai, Japan*
[3]*Department of Physics, Freie Universität Berlin, 14195 Berlin, Germany*

*Authors to whom correspondence should be addressed: stephane.mangin@univ-lorraine.fr

**Abstract**

The discovery of all-optical ultra-fast deterministic magnetization switching has opened up new possibilities for manipulating magnetization in devices using femtosecond laser pulses. Previous studies on single pulse all-optical helicity-independent switching (AO-HIS) have mainly focused on perpendicularly magnetized thin films. This work presents a comprehensive study on AO-HIS for in-plane magnetized $Gd_xCo_{100-x}$ thin films. Deterministic single femtosecond laser pulse toggle magnetization switching is demonstrated in a wider concentration range (x=10% to 25%) compared to the perpendicularly magnetized counterparts with GdCo thicknesses up to 30 nm. The switching time strongly depends on the $Gd_xCo_{100-x}$ concentration, with lower Gd concentration exhibiting shorter switching times (less than 500 fs). Our findings in this geometry provide insights into the underlying mechanisms governing single pulse AO-HIS, which challenge existing theoretical predictions. Moreover, in-plane magnetized $Gd_xCo_{100-x}$ thin films offer extended potential for opto-spintronic applications compared to their perpendicular magnetized counterparts.

# I. INTRODUCTION

The advancement of magnetic data storage, memories, and logic necessitates a fast and energy-efficient method for manipulating magnetization in thin magnetic media and heterostructures like magnetic tunnel junctions and spin valves. While significant progress has been made in current-induced magnetic switching over the past 25 years, the typical time required for this process is still orders of magnitude slower than optically induced magnetization manipulation [1-4]. In 2012, Ostler *et al.* achieved field-free ultrafast magnetization reversal by irradiating a femtosecond laser pulse onto a ferrimagnetic GdFeCo alloy [5]. This breakthrough paved the way for All-Optical Helicity-Independent Switching (AO-HIS), although its underlying mechanism remains a topic of debate [6-11]. The reversal mechanism is primarily attributed to a pure ultrafast thermal effect on magnetization, enabled by ultrafast demagnetization and subsequent angular momentum exchange between rare-earth and transition-metal sublattices on sub-picosecond and picosecond timescales [5,10-16]. This enables reliable writing of magnetic bits at GHz frequencies [17,18].

The AO-HIS has been mainly reported in perpendicularly magnetized Gd-based alloys or multilayers [5-23]. GdFeCo and GdCo alloys have been extensively studied. They are ferrimagnetic alloys for which the magnetization of the Gd sublattice ($M_{Gd}$) is exchange-coupled antiferromagnetically to the magnetization of the Transition Metal (Fe, Co) sublattice ($M_{Co}$). The resulting magnetization depends on the alloy concentration and temperature. Due to the antiferromagnetic coupling, the net magnetization is zero at the so-called compensation composition ($x_{comp}$) which depends on temperature. It has been shown that AO-HIS, measured at room temperature, occurs in $Gd_x(FeCo)_{100-x}$ only when x is close to $x_{comp}$ (at room temperature) within a few percent [10,11,19-22]. Theoretical models tend to confirm that only alloys with a concentration close to the $x_{comp}$ could exhibit AO-HIS [21]. In 2015, Atxitia *et al.* used an atomistic stochastic Landau-Lifshitz-Gilbert equation for semi-classical spins, described by a Heisenberg Hamiltonian, to model AO-HIS in rare earth-transition metal ferrimagnetic alloys and concluded that a low net magnetization is an important ingredient for an energy-efficient AO-HIS [24]. The same conclusions were drawn a few years later by Jakobs *et al.* [11] by using an atomistic model and the so-called two-temperature model, and by Davies *et al.* [22] who used a phenomenological framework showing theoretical agreement with experimental results.

With the exception of a single experimental study conducted on in-plane magnetized $Gd_{25}(FeCo)_{75}$ microstructures with a specific Gd concentration [5], all previous experimental results in this field have been obtained using samples exhibiting strong perpendicular magnetic anisotropy (PMA) [5-22]. Undeniably, PMA samples offer technical advantages for performing experiments, as the pump and probe beams can be directed perpendicular to the sample's surface. However, these samples may face challenges related to the stabilization of maze domain structures, which can limit the observation of AO-HIS only when the net magnetization and the thickness of the layers are low [20,23-26]. In cases where the magnetization is high and the alloy concentration is far from $x_{comp}$, the magnetic configuration tends to break into domains after laser excitation to minimize dipolar (demagnetization) energy [25]. Consequently, we can question the theoretical notion that a concentration close to compensation (i.e., low magnetization) is an intrinsic and mandatory requirement for achieving AO-HIS or whether it is an extrinsic effect driven by domain structure stabilization. In contrast to PMA systems, in-plane magnetized thin films are not expected to experience a strong dipolar field that induces a multidomain state. Therefore, studying these films can help addressing uncertainties regarding intrinsic and extrinsic effects.

This manuscript presents a comprehensive investigation into the manipulation of magnetization using a single femtosecond laser pulse in in-plane magnetized $Gd_xCo_{100-x}$ thin films. Unlike perpendicularly magnetized $Gd_x(FeCo)_{100-x}$ films, we were able to successfully demonstrate All-Optical Helicity-Independent Switching (AO-HIS) over a wide range of concentrations (5% < x < 30%) and thicknesses (5 to 30 nm). This observation challenges existing theoretical predictions, which suggested a narrow concentration range around compensation for AO-HIS. By analyzing the laser-induced effects, we identified three distinct threshold fluences that play a crucial role in observing AO-HIS: the fluence required for magnetization switching ($F^{th}_{Sw}$), the fluence for demagnetization ($F^{th}_{Dem}$) for which the multidomain patterns are created due to excessive heating of the sample, and the fluence that causes irreversible changes or damage to the magnetic properties ($F^{th}_{Dam}$). $F^{th}_{Sw}$ exhibited a minimum at compensation, while $F^{th}_{Dem}$ increased with the sample's Curie temperature. The magnetization dynamics during the reversal process resembled those observed in perpendicularly magnetized films, with the fastest switching occurring for the lowest Gd concentration.

## II. EXPERIMENTAL RESULTS

### A. The magnetic properties of in-plane magnetized $Gd_xCo_{100-x}$ thin films

$Gd_xCo_{100-x}$ ferrimagnetic thin films consisting of Glass/Ta (3 nm)/$Gd_xCo_{100-x}$ (t nm)/Cu (1 nm)/Pt (3 nm) were prepared with a wide range of Gd concentration, x, varying from 5% to 35% and a wide range of $Gd_xCo_{100-x}$ thickness, t, varied from 5 to 35 nm. In all cases, the $Gd_xCo_{100-x}$ interfaces have been designed to keep an easy uniaxial anisotropy axis in-plane for any x and t values. Fig. 1(b) summarizes the evolution of the magnetization and coercivities of the sample as a function of the concentration for t=5 nm. As expected, the coercivity diverges, and the net magnetization reaches zero around x=20%, corresponding to the compensation composition ($x_{comp}$) at room temperature. $M_{Co}$ is dominant over $M_{Gd}$ when x<20%, while $M_{Gd}$ becomes dominant as x>20%. As shown in Supplemental Material 1, all hysteresis loops measured with an in-plane field have a remanence close to one, revealing the existence of a well-defined in-plane magnetic anisotropy axis in all $Gd_xCo_{100-x}$ films. This is also confirmed by the angle-dependent remanence measurements shown in Supplemental Material 9. Moreover, the out-of-plane magnetic field-dependent saturation magnetization measurements for all the samples are provided in Supplemental Material 2, indicating that the hard axis is mainly perpendicular to the sample surface. The evolution of Kerr microscope images as a function of an in-plane magnetic field applied along the easy axis tends to show that the domain size and the domain growth by domain wall motion are similar for all GdCo concentration (Supplemental Material 3).

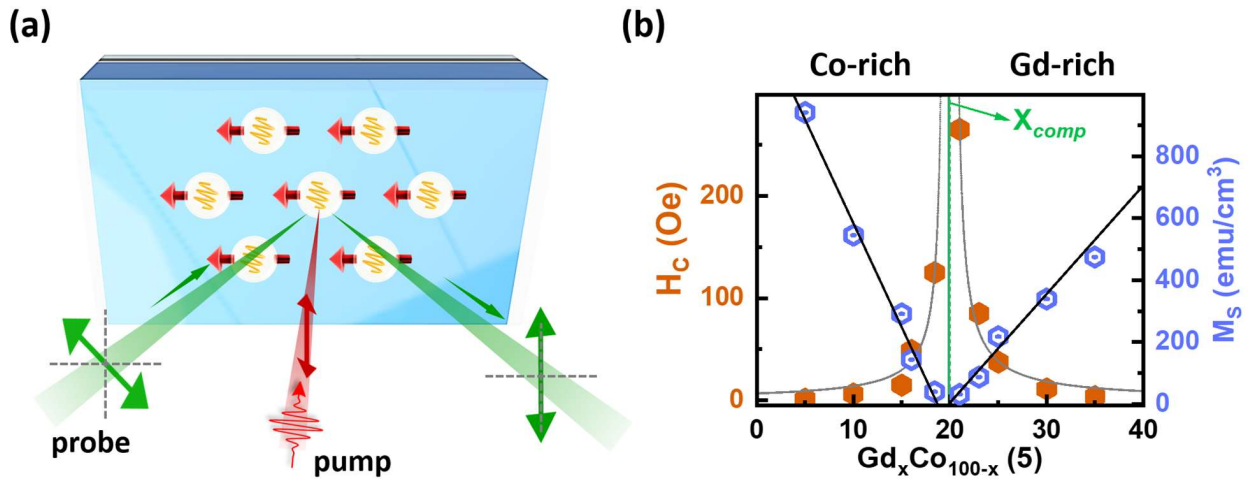

**FIG. 1.** (a) Schematic of the static single laser pulse and time-resolved measurements based on longitudinal magneto-optic Kerr effect (MOKE). The linearly polarized pump laser pulse (800 nm, 150 fs) is shined perpendicular to the film plane, whereas the linearly p-polarized probe beam (515 nm) angle of incidence is 45°. The reflected optical probe beam is sent to a camera for MOKE images. The sample is magnetized in-plane using an external in-plane external magnetic field before pumping the laser pulse. The red three-dimensional arrows represent the Co magnetic moments. Experiments were carried out at room temperature. (b) Variation of the coercive field $H_C$ (solid orange symbols) and the saturation magnetization $M_S$ (open purple symbols) as a function of Gd content x in Glass/Ta (3 nm)/$Gd_xCo_{100-x}$ (5 nm)/Cu (1 nm)/Pt (3 nm).

**B. Single pulse all-optical helicity-independent switching for in-plane magnetized $Gd_xCo_{100-x}$ alloys**

A sketch of the experimental setup is shown in Fig. 1(a). It is based on a standard longitudinal MOKE configuration that allows tracking the in-plane magnetization changes after shining linearly polarized femtosecond laser pulses. One important parameter, the pulse length, has been fixed to 150 fs, and further details are given in the experimental section. Fig. 2(a) demonstrates that for a $Gd_{15}Co_{85}$ in-plane magnetized thin film, it is possible to observe a deterministic single pulse all-optical helicity-independent switching (AO-HIS), independently of the initial direction of magnetization. First, two domains with opposite directions along the in-plane easy axis were created using an in-plane external magnetic field; then, laser pulses were shined at three different positions under zero applied field. As a result, one can clearly observe that a full magnetization switching is observed for the two initial magnetization directions. Moreover, we confirmed that a full switching is indeed observed since the contrast variations between the magnetic field-induced and the laser-induced switching are the same.

The laser-induced magnetization switching obtained for various $Gd_xCo_{100-x}$ concentrations is shown in Figures 2(b) and (c). They present the MOKE images obtained after shining 0, 1, 2, and 3 pulses at the same position for a series of 5 nm $Gd_xCo_{100-x}$ alloy samples. AO-HIS is demonstrated for x ranging from 10% to 25%. In this case, a fully reversed domain appears after the first pulse, completely vanishes after the second pulse, and fully re-appears after the third one indicating a perfect toggle-switching behavior. In addition, a multidomain state is formed in the center region of the spot when the laser fluence is large enough (Supplemental Material 4). Perfect toggle switching for in-plane magnetized GdCo films is observed even after 1000 pulses, as shown in Supplemental Material 5, demonstrating the endurance of AO-HIS in such films potentially for technological applications. To make sure that dipolar fields are not affecting the switching, AO-HIS has been demonstrated when shining laser pulses on the boundary between two domains, as shown in Supplemental Material 6. The effect of single laser pulses was also carried out for samples with excessive content of Co (i.e., x=5%) and Gd (i.e., x=30%). For x=5%, neither a typical round-shaped switching pattern nor a multidomain state was observed before the degradation of the sample, as shown in Supplemental Material 7. For x=30%, either no switching or a multidomain state was obtained (Supplemental Material 8). Nevertheless, as shown for heat-assisted magnetic recording, those disordered patterns can be removed, and magnetization can be reversed by applying a tiny external field along the direction opposite to the initial magnetization direction [27,28]. Supplemental Material 9 compares the effect of light when the sample is saturated along the easy and in-plane hard axes for $Gd_{15}Co_{85}$. Along the easy axis, as reported earlier, full switching is demonstrated. However, when the sample is first saturated along the in-plane hard axis, and then the field is removed, the remanent state is multidomain; consequently, no single-domain all-optical switching could be observed; however, partial light-induced switching is clearly demonstrated.

To summarize AO-HIS in GdCo in-plane magnetized samples, Fig. 2(d) shows the threshold fluence needed to observe switching ($F^{th}_{Sw}$) and threshold fluence needed to demagnetize the sample ($F^{th}_{Dem}$; i.e., the multidomain state is formed in the irradiation area in the case of too high pump pulse fluence, as shown in Fig. S4(c)) as a function of the Gd concentrations. This diagram clearly demonstrates that the window of Gd concentration (Δx) showing the AO-HIS is much larger for in-plane magnetized GdCo alloys (Δx~20%) compared to perpendicular magnetic anisotropy (PMA) GdFeCo alloys (Δx~5%) [10,19,20] and pure PMA GdCo alloys (Δx~9%) [21,29]. The diagram also shows that a minimum in $F^{th}_{Sw}$ can be observed close to the $x_{comp}$, whereas $F^{th}_{Dem}$ decreases monotonically with increasing Gd concentration, following the Curie temperature ($T_C$) (Supplemental Material 10). Note that a third threshold fluence ($F^{th}_{Dam}$) should be defined as the fluence threshold which damages the sample such that it cannot recover its initial magnetic state. The reason why no AO-HIS is observed for x= 5% and x=30% would then come from two different reasons. As we will discuss later, no AO-HIS is observed because $F^{th}_{Sw} > F^{th}_{Dam}$ for x= 5%, and because $F^{th}_{Sw} > F^{th}_{Dem}$ for x=30%. For the PMA $Gd_x(FeCo)_{100-x}$ alloys, it has been shown that the laser pump energy required to produce AO-HIS is minimum around $x_{comp}$ [19,21,30], which is in agreement with in-plane magnetized GdCo alloys, hinting that in-plane magnetized GdCo alloys should share the same switching mechanism as perpendicularly magnetized GdCo alloys.

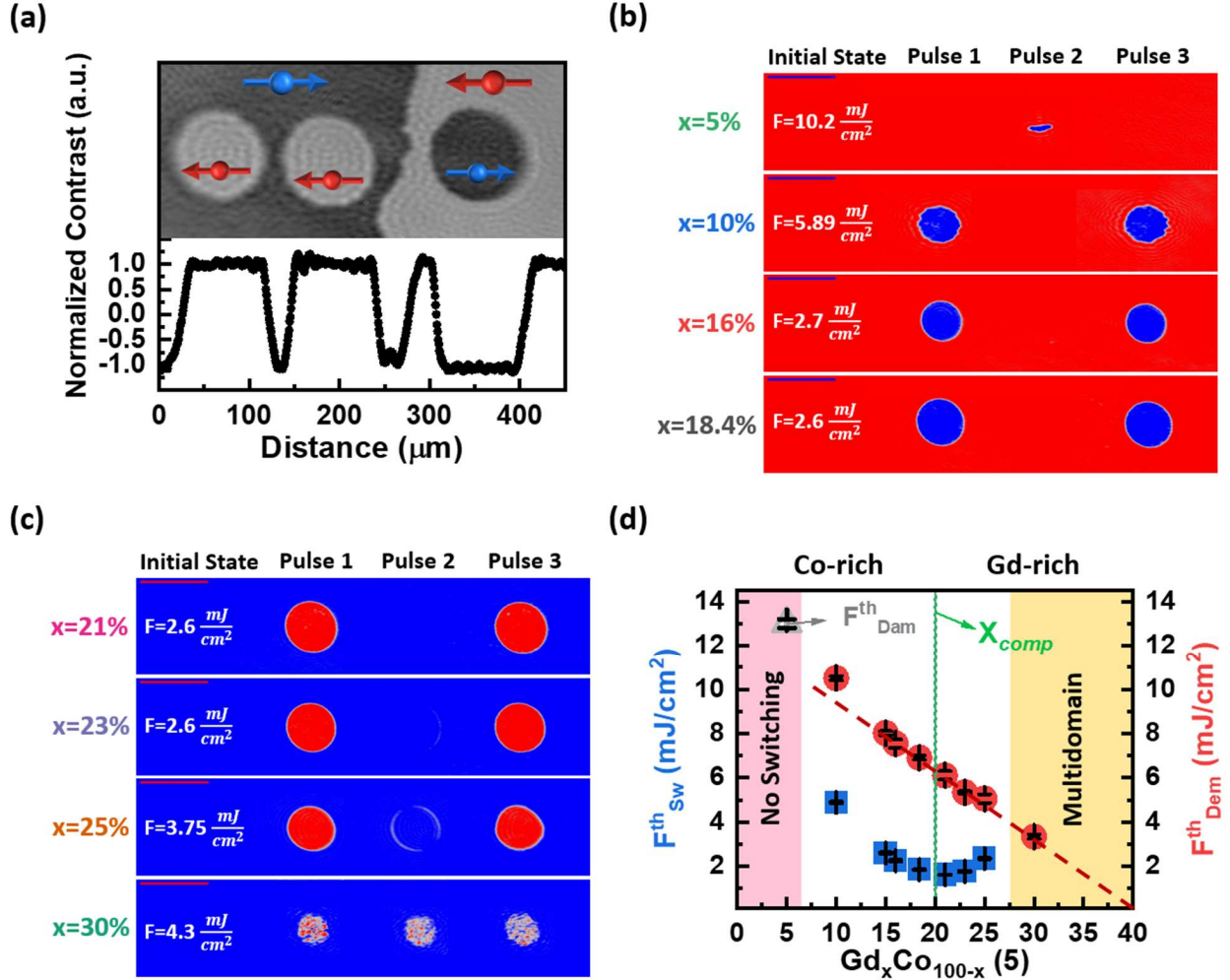

**FIG. 2. AO-HIS in 5 nm thick $Gd_xCo_{100-x}$ alloys** (a) Kerr images and normalized contrast cross-section after three single laser shots inducing AO-HIS for a $Gd_{15}Co_{85}$ thin film starting from a two-domain magnetic state of opposite directions along the easy axis. The blue and red arrows indicate the Co sublattice's magnetization direction. Magneto-optic contrast obtained after a single 150 fs laser pump pulse on various in-plane magnetized $Gd_xCo_{100-x}$ alloys: (b) for Co-dominant samples and (c) for Gd-dominant samples. For each measurement, the laser pulses were shined at the same position. A scale bar of 100 μm is presented. (d) The switching threshold ($F^{th}_{Sw}$) and demagnetization threshold ($F^{th}_{Dem}$) fluences as a function of Gd concentrations. Note that the grey open triangle indicates the threshold to permanently damage the sample ($F^{th}_{Dam}$), and the red dashed line guides the eye for the $F^{th}_{Dem}$.

## C. Thickness-dependent single pulse all-optical helicity-independent switching for in-plane magnetized GdCo alloys

After investigating the influence of the GdCo alloy concentration on AO-HIS, we now study its thickness dependence. For this study, we fixed the arbitrary Gd concentration x to 25% ($Gd_{25}Co_{75}$). The samples normalized longitudinal MOKE hysteresis loops for various thicknesses are shown in Supplemental Material 11. Square hysteresis loops are observed along the in-plane easy axis for all thicknesses ranging from 5 to 35 nm, demonstrating a strong in-plane anisotropy easy axis for all thicknesses. Fig. 3(a) demonstrates deterministic toggling AO-HIS for thickness (t) ≤30 nm. $F^{th}_{Sw}$ and $F^{th}_{Dem}$, as a function of the $Gd_{25}Co_{75}$ thicknesses, are shown in Fig. 3(b).

Both $F^{th}_{Sw}$ and $F^{th}_{Dem}$ are following the same trend. For low GdCo thickness (t<20 nm), the two threshold fluences increase linearly with thickness, which can be understood by supposing that the amount of energy absorbed by the GdCo layer is depth independent in this thickness range. Switching and demagnetization depend on the deposited energy density and, consequently, on temperature. The fact that the two fluences tend to diverge with the GdCo thickness implies that the absorption can no longer be considered uniform. Indeed, the depth-dependent absorption profiles obtained using the transfer matrix method reflect the heat absorption at different film depths (Supplemental Material 12) [31,32]. The obtained absorption gradient implies that the laser pump pulse heats the system inhomogeneously, indicating that the front part of the sample (facing the laser pump pulse) absorbs more laser energy than the latter part of the magnetic layer. If the entire GdCo layer needs to reach a certain temperature before switching, the laser fluence will then increase nonlinearly and tend to diverge as the sample thickness increases. We can then speculate that the speckled magnetic domain observed for $Gd_{25}Co_{75}$ with a thickness of t=35 nm could be understood by the fact that the front part of the sample absorbed enough energy to demagnetize this area, whereas the back part does not absorb enough energy to switch, resulting in a maze domain structure.

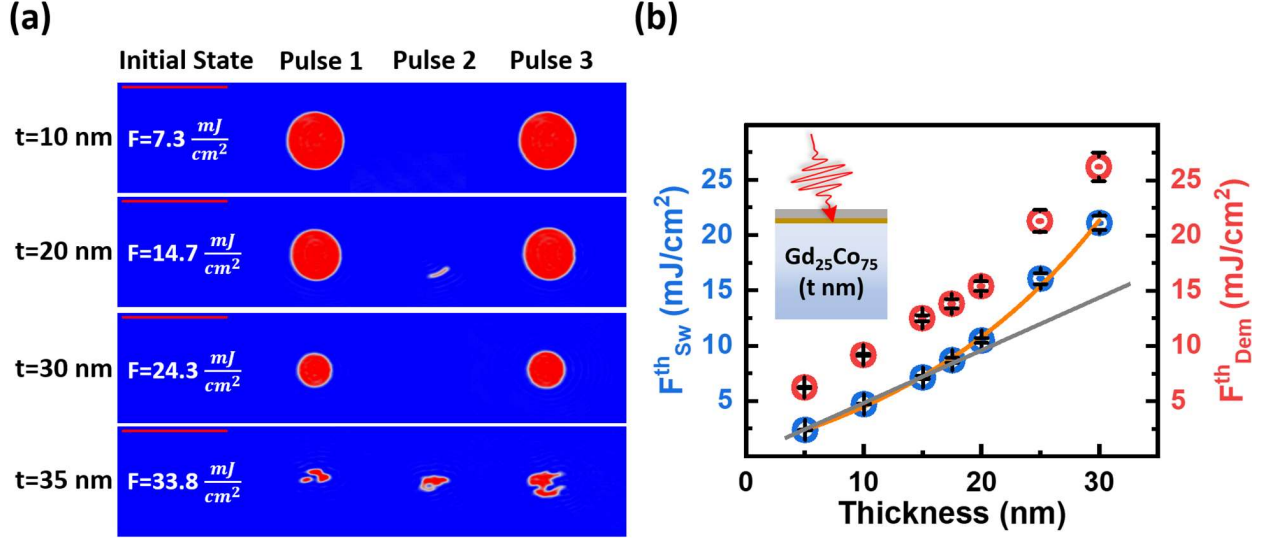

**FIG. 3.** (a) Static MOKE images obtained after single laser pump pulse irradiation for different thicknesses (t) of the in-plane magnetized $Gd_{25}Co_{75}$ thin film. The scale bar is 100 μm long. (b) Dependence of $F^{th}_{Sw}$ and $F^{th}_{Dem}$ as a function of thickness. The orange line is obtained from a fitting using an exponential function. The linear grey line is a guide for the eye.

## D. Time-resolved magneto-optic Kerr measurements

After investigating the effect of a single femtosecond laser pulse using Kerr images which are taken several seconds after the excitation, we are now probing the fast magnetization dynamics of 5 nm $Gd_xCo_{100-x}$ samples using longitudinal time-resolved MOKE (TR-MOKE). TR-MOKE measurements were performed for samples showing toggle-switching behavior. Since the magnetization dynamics depend strongly on the laser fluence, in an effort to normalize the laser fluence, for each concentration, we used a laser fluence 1.2 larger than the previously determined threshold fluence $F^{th}_{Sw}$. In Figures 4(a) and (b), all curves showed similar features: a fast initial drop followed by a slower decrease and a saturation toward -1. This type of feature has already been seen for the AO-HIS in the single-layer PMA GdFeCo alloys demonstrated by different research groups [6,8,10,11,21]. This indicates, as expected at short time scales, that the AO-HIS effect is independent of the magnetic anisotropy and should originate from a purely ultrafast thermal effect [5,6,10,11,21]. To qualitatively describe the dynamics traces, we defined two characteristic times: T1 corresponding to the time at the end of the first fast drop and T2 to the time corresponding to the end of the slower decrease, as shown in Supplemental Material 13. In Fig. 4(c), T1 and T2 are plotted as a function of the Gd concentration. Those measurements show that the magnetization dynamics of Co slow down with increasing Gd concentration.

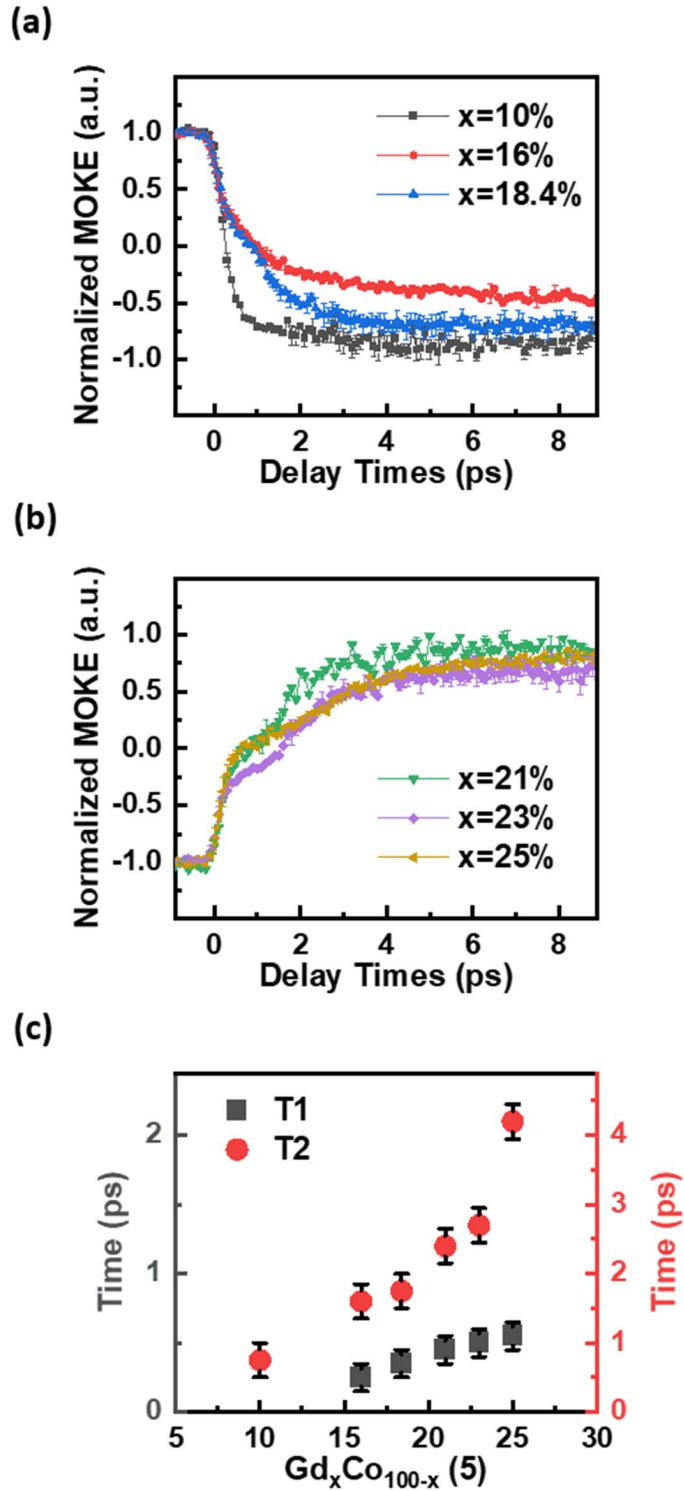

**FIG. 4.** Normalized longitudinal MOKE signal as a function of time for $Gd_xCo_{100-x}$ (5 nm) for (a) Co-dominant alloys (x=10%, 16%, and 18.4%) and (b) Gd-dominant alloys GdCo (x=21%, 23%, and 25%). The measurements were obtained with an external magnetic field along the in-plane initial easy-axis direction and for fluences 1.2 times larger than the fluence threshold ($F^{th}_{Sw}$). (c) Evolution of two characteristic times (T1 and T2) of the switching as a function of x the $Gd_xCo_{100-x}$ concentration.

### III. DISCUSSION

The obtained results provide clear evidence of *single pulse*, *ultra-fast* All-Optical Helicity-Independent Switching (AO-HIS) in in-plane magnetized GdCo alloys. Notably, this AO-HIS phenomenon is observed over a significantly larger concentration and thickness range compared to previous experimental observations in perpendicularly magnetized systems [10,11,19,20,29,33] and theoretical predictions [10,11,21,24]. The extended concentration range can be attributed to the absence of a stabilized multidomain state due to dipolar fields (stray fields or demagnetization fields) far from magnetization compensation composition ($x_{comp}$). In perpendicularly magnetized samples, the generation of a multidomain state by dipolar fields requires the definition of a threshold fluence, denoted as $F^{th}_{Multi}$ [25]. Zhang *et al.* demonstrated that a criterion for observing AO-HIS is that $F^{th}_{Sw}$ (the fluence needed for magnetization switching) is less than $F^{th}_{Multi}$. Otherwise, the switching process is overshadowed by the formation of the multidomain state, which occurs at a longer timescale. As dipolar fields increase with sample magnetization and thickness, $F^{th}_{Multi}$ decreases when moving away from $x_{comp}$ and for thicker samples.

Because $F^{th}_{Multi}$ does not need to be considered in the in-plane magnets, the observation of AO-HIS is not limited to low-thickness or low-magnetization alloys (close to compensation in the case of ferrimagnetic alloys). Accordingly, the critical criteria become $F^{th}_{Sw} < F^{th}_{Dam}$ and $F^{th}_{Sw} < F^{th}_{Dem}$. We can suppose that $F^{th}_{Dam}$ does not depend on the Gd concentration, whereas $F^{th}_{Dem}$ decreases with Gd concentration as a consequence of the decrease of Curie temperature ($T_C$) [34]. The latter is clearly observed in Fig. 2(d), where $F^{th}_{Dem}$ decreases almost linearly with increasing concentration of Gd. If we extrapolate the evolution of $F^{th}_{Dem}$ as a function of Gd concentration, $F^{th}_{Dem}$=0 can be reached for x=40%, the concentration at which the $T_C$ approaches 300 K [34,35]. For x=30% (highest Gd concentration in this work), the AO-HIS cannot be seen because of $F^{th}_{Sw}$ > $F^{th}_{Dem}$. On the other hand, as detailed in the work of Zhang *et al.*, $F^{th}_{Sw}$ depends both on the alloy's $T_C$ and the amount of angular momentum generated [25]. In our work, for x=5% (lowest Gd concentration in this work), the AO-HIS cannot be observed due to the fact that $F^{th}_{Sw}$ > $F^{th}_{Dam}$. Because of the high $T_C$ value of the alloy for low Gd concentration, a laser fluence larger than $F^{th}_{Dam}$ would be needed to switch the magnetization. Furthermore, a minimum $F^{th}_{Sw}$ appears close to $x_{comp}$ in our study, as observed in PMA GdFeCo alloys [11,19,21,24]. This could be in accord with Barker *et al.* who suggested that a nonequilibrium energy transfer between the ferromagnetic- and antiferromagnetic-like magnon branches is maximized near $x_{comp}$, resulting in a minimum $F^{th}_{Sw}$ around $x_{comp}$ [30].

Finally, the fastest switching is obtained for low Gd concentration and tends to increase as the Gd concentration increases, as shown in Fig. 4(c). We confirm here that the Co spin-lattice coupling is an efficient channel for angular momentum dissipation in ferrimagnetic alloys [36]. Indeed, it has been demonstrated that for GdCo alloy, the angular momentum dissipation into the lattice for Gd is less than for Co [6,36]. Adding Gd leads to changes in the speeds and amplitudes of both sublattices' demagnetization [21,37]. As confirmed experimentally, when ultrafast demagnetization occurs, the angular momentum will either dissipate locally into the lattice or be transferred as a spin current [4,36-42]. Therefore, the more Gd is introduced, the less dissipation channels are available at a short time scale, resulting in the slowdown of the overall magnetization dynamics.

**IV. CONCLUSION**

In conclusion, our systematic study has successfully demonstrated deterministic All-Optical Helicity-Independent Switching (AO-HIS) using a femtosecond laser pulse for in-plane magnetized $Gd_xCo_{100-x}$ thin films at room temperature. The observation of ultrafast switching across a wide concentration range, in contrast to perpendicularly magnetized counterparts, is due to the absence of perpendicular demagnetization fields which tend to break magnetization into domains. These findings challenge the notion that compensation composition and temperature are necessary to ensure AO-HIS. However, it is crucial to create conditions for which the switching fluence threshold is lower than the thresholds leading to sample damage and demagnetization in order to observe the desired switching behavior. Additionally, our results indicate that switching can be achieved in relatively thick films, depending on the demagnetization process, providing valuable insights for future stack engineering and optimization of AO-HIS. Furthermore, the magnetization dynamics observed during the reversal of in-plane magnetized GdCo alloys are similar to those observed in perpendicularly magnetized counterparts. Notably, the demagnetization speed and subsequent switching slow down with increasing Gd concentration. These experimental findings give new insights into the magnetization reversal of Gd-based materials, which hold significant implications for the development of future ultrafast spintronic memory devices.

## V. EXPERIMENTAL SECTION/METHODS

### A. Sample preparations

Glass/Ta (3 nm)/$Gd_xCo_{100-x}$ alloys (t nm)/Cu (1 nm)/Pt (3 nm) with different x and t values were prepared through magnetron co-sputtering with elemental targets processed under an Argon gas pressure of approximately $10^{-3}$ mbar. Multilayered typical structures were deposited onto 15×10-mm glass substrates. During the room-temperature sample deposition, the sample holder was fixed in a specific direction without rotating the holder, which produced a well-defined in-plane anisotropy in $Gd_xCo_{100-x}$ (t nm), as shown in Fig. S9(b). In samples for the thickness-dependent all-optical helicity-independent switching (AO-HIS) experiments, the $Gd_xCo_{100-x}$ (t nm) concentration is fixed at x=25% while t varies from 5 nm to 35 nm with an interval of 5 nm.

### B. Characterizations

Static single pulse and time-resolved measurements were performed in a longitudinal magneto-optic Kerr effect (MOKE) configuration. Samples were housed in an optical setup integrated with a dipole electromagnet, allowing for a variable field (0–0.3 T) operation along the sample plane. The linearly polarized pump pulse with a spot size diameter of ~120 μm was normal incident to the sample surface. On the other hand, the linearly p-polarized probe beam with low energy was used to impinge on the sample at an incidence angle of 45 degrees to obtain the MOKE images using a complementary metal oxide semiconductor camera. Both laser beams were shot on the sample side to perform all the measurements. In this work, the wavelength of the pump pulse was fixed at 800 nm, and the one for the probe beam was set at 515 nm, which mainly reflects the magnetic signals of the Co sublattice of GdCo alloys. One crucial parameter for all the AO-HIS measurements: the laser pulse length was fixed at 150 fs. All the measurements were performed at ambient temperature.

On the one hand, in the static single pulse AO-HIS measurements, the external in-plane magnetic field (oriented parallel to the substrate and along the in-plane easy axis, as indicated in Fig. S9(c)) with a strength larger than the sample coercivity was first applied to initialize the sample. Thereby all the moments in the alloy were aligned in the direction of the in-plane easy axis under a zero field. Subsequently, the samples were irradiated by different numbers of laser pump pulses without any external magnetic field to perform the static single pulse AO-HIS measurements. The repetition rate of the laser pump pulse was set to 100 kHz for all static single pulse AO-HIS measurements.

On the other hand, in the time-resolved MOKE imaging measurements, an in-plane magnetic field with a strength around the sample coercivity was always applied along the initial in-plane easy axis direction of magnetization to initialize the sample before each pump pulse. The value varies from one sample to another due to the different coercivities of all investigated samples. Moreover, we used the MOKE imaging configuration to perform the dynamics measurements, which may include some artificial optical signals originating from a substrate or environment. To remove those signals, we took the intensity difference for opposite magnetization directions at negative time delay to normalize the presented data, as described in a previous study [42]. Here, the time zero (time delay=0) was determined by the time delays where the derivative of the magnetization dynamics trace is maximal. Since the threshold to permanently damage the sample is different from one sample to another, the endurance capacity of the temperature of the sample within a given time is also different. In this study, for x=16%, 18.4%, 21%, and 23%, a repetition rate of 100 kHz was used for both the pump and probe beam, while for x=10% and 25%, the value was reduced to 10 kHz.

The saturation magnetization and the coercivity in Fig. 1(b) and the magnetization versus magnetic field curves in Fig. S2 were examined using a superconducting quantum interference device.

**ACKNOWLEDGMENT**

The authors thank Eric Fullerton and Bert Koopmaans for fruitful discussion. This work is supported by the ANR-20-CE09-0013 UFO, the Institute Carnot ICEEL for the project "CAPMAT" and FASTNESS, the Région Grand Est, the Metropole Grand Nancy for the Chaire PLUS, the interdisciplinary project LUE "MAT-PULSE", part of the French PIA project "Lorraine Université d'Excellence" reference ANR-15-IDEX-04-LUE, the "FEDERFSE Lorraine et Massif Vosges 2014-2020" for the project PLUS and IOMA, a European Union Program, the European Union's Horizon 2020 research and innovation program COMRAD under the Marie Skłodowska-Curie grant agreement No 861300, the ANR project ANR-20-CE24-0003 SPOTZ,. This article is based upon work from COST Action CA17123 MAGNETOFON, supported by COST (European Cooperation in Science and Technology). This work was supported by the French National Research Agency through the France 2030 government grants EMCOM (ANR-22-PEEL-0009). All fundings were shared equally among all authors.

# Single laser pulse induced magnetization switching in in-plane magnetized GdCo alloys

Jun-Xiao Lin,[1] Michel Hehn,[1,2] Thomas Hauet,[1] Yi Peng,[1] Junta Igarashi,[1] Yann Le Guen,[1] Quentin Remy,[3] Jon Gorchon,[1] Gregory Malinowski,[1] Stéphane Mangin,*[1,2] and Julius Hohlfeld[1]

[1]*Université de Lorraine, CNRS, Institut Jean Lamour, F-54000 Nancy, France*
[2]*Center for Science and Innovation in Spintronics, Tohoku University, Sendai, Japan*
[3]*Department of Physics, Freie Universität Berlin, 14195 Berlin, Germany*

*Authors to whom correspondence should be addressed: stephane.mangin@univ-lorraine.fr

## S1. Longitudinal magneto-optic Kerr effect (MOKE) hysteresis loops of in-plane magnetized $Gd_xCo_{100-x}$ films at room temperature

The longitudinal MOKE hysteresis loops were obtained by applying a magnetic field along the in-plane easy axis. Since MOKE is mainly sensitive to the Co sublattice, the sign of the hysteresis loops depends on the Gd concentration x: counterclockwise (respectively clockwise) hysteresis loops are obtained when the magnetization of the Co sublattice, $M_{Co}$, is higher (respectively lower) than the magnetization of the Gd sublattice, $M_{Gd}$. In Fig. S1, we can clearly see that $M_{Co}$ is dominant over $M_{Gd}$ when x<20%, while $M_{Gd}$ becomes dominant when x>20%, which is consistent with the results observed in Fig. 1(b). The compensation composition is then $x_{comp} \sim 20\%$. In all cases, the hysteresis loops have a remanence close to one, revealing the existence of a well-defined in-plane magnetic anisotropy axis in all the studied $Gd_xCo_{100-x}$ films.

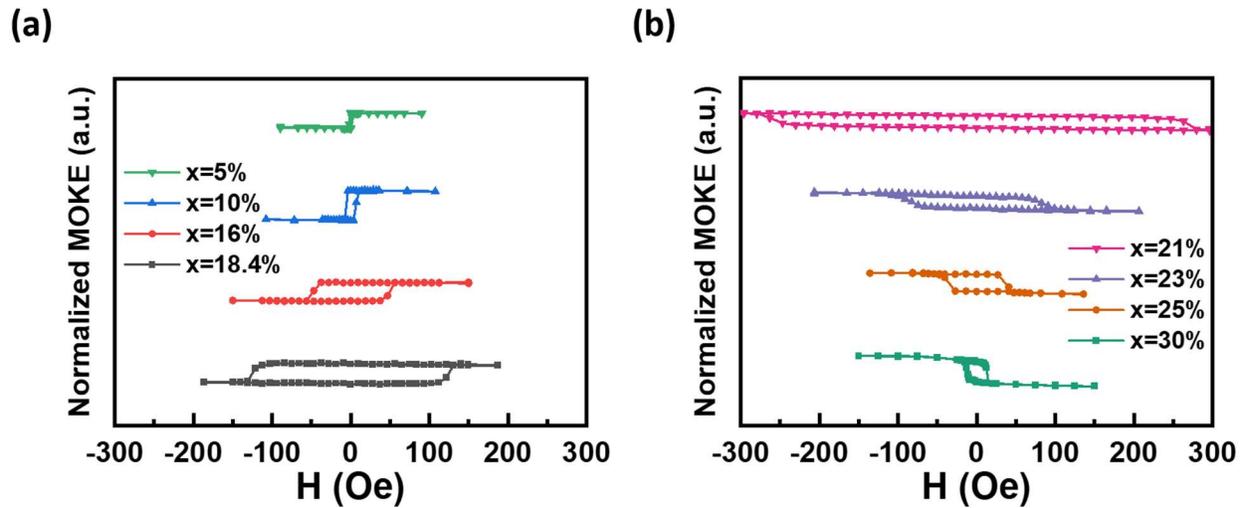

**FIG. S1.** Normalized MOKE signal as a function of magnetic field (H) applied along the in-plane anisotropy axis for various $Gd_xCo_{100-x}$ films concentration (a) for Co-rich alloys (x=5%, 10%, 16%, and 18.4 %) and (b) for Gd-rich alloys (x=21%, 23%, 25%, and 30 %).

## S2. Magnetic field-dependent magnetization (M-H) measurements of $Gd_xCo_{100-x}$ films at room temperature

Magnetization versus magnetic field curves, with the external magnetic field applied along the out-of-plane direction, are shown in Fig. S2 for the studied samples. Those results indicate that the out-of-plane direction is a hard axis.

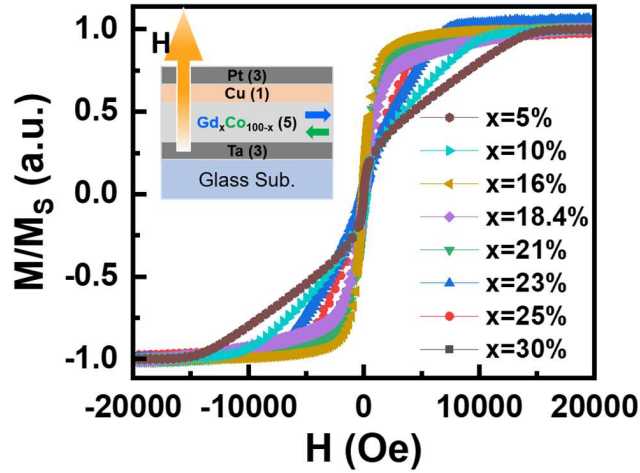

**FIG. S2.** Normalized magnetization ($M/M_S$) as a function of magnetic field (H) applied along the direction perpendicular to the film plane for various Glass/Ta (3 nm)/$Gd_xCo_{100-x}$ (5 nm)/Cu (1 nm)/Pt (3 nm).

## S3. Magnetic domain imaging of in-plane magnetized Gd$_x$Co$_{100-x}$ thin films

Initially, the sample was saturated using a strong external in-plane magnetic field, then the field (with a direction opposite to the saturating field) was reduced close to coercivity to generate magnetic domains. The obtained Kerr microscopy images suggest that the characteristic domain size and shape are similar for all the studied Gd$_x$Co$_{100-x}$ films. The typical domain size is comparable to the laser pump spot size, which is around 120 μm in diameter.

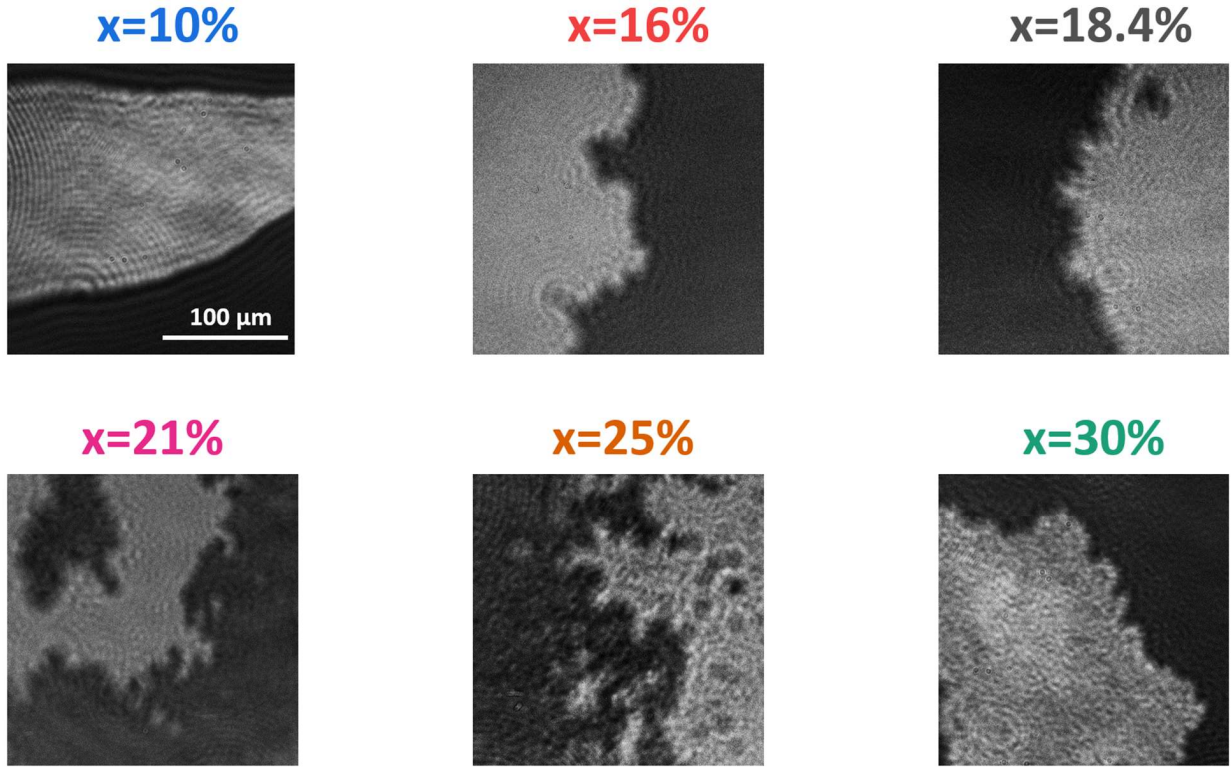

**FIG. S3.** Kerr microscope images obtained on Glass/Ta (3 nm)/Gd$_x$Co$_{100-x}$ (5 nm)/Cu (1 nm)/Pt (3 nm) under an in-plane field close to coercivity.

## S4. Reversed domain diameter versus laser pump pulse energy

The sample is excited with a single laser pump pulse with various energies. Fig. S4(a) shows that the diameter of the written domain increases as the pumped energy increases. We consider that the laser beam has a Gaussian intensity profile with a maximum value at the center of the beam. When the energy is too high, a demagnetization state is observed in the center of the circular reversed domain. The threshold pump pulse energy ($E_{th}$) to observe all-optical helicity-independent switching (AO-HIS) can be determined from $d = D\sqrt{\frac{E}{E_{th}}}$, where the $d$ (the diameter of the reversed domain) and $E$ (the laser pump pulse energy) are the controlled experimental values, as shown in Fig. S4(b). Then, by fitting the set of data points, the $D$ (laser-spot diameter) and the $E_{th}$ (threshold pump pulse energy) can be evaluated. By knowing $E_{th}$ and $D$, the threshold fluence ($F_{th}$) is given by $\frac{E_{th}}{\pi \cdot (\frac{D}{2})^2}$. Detailed information can be found in a previous work [1].

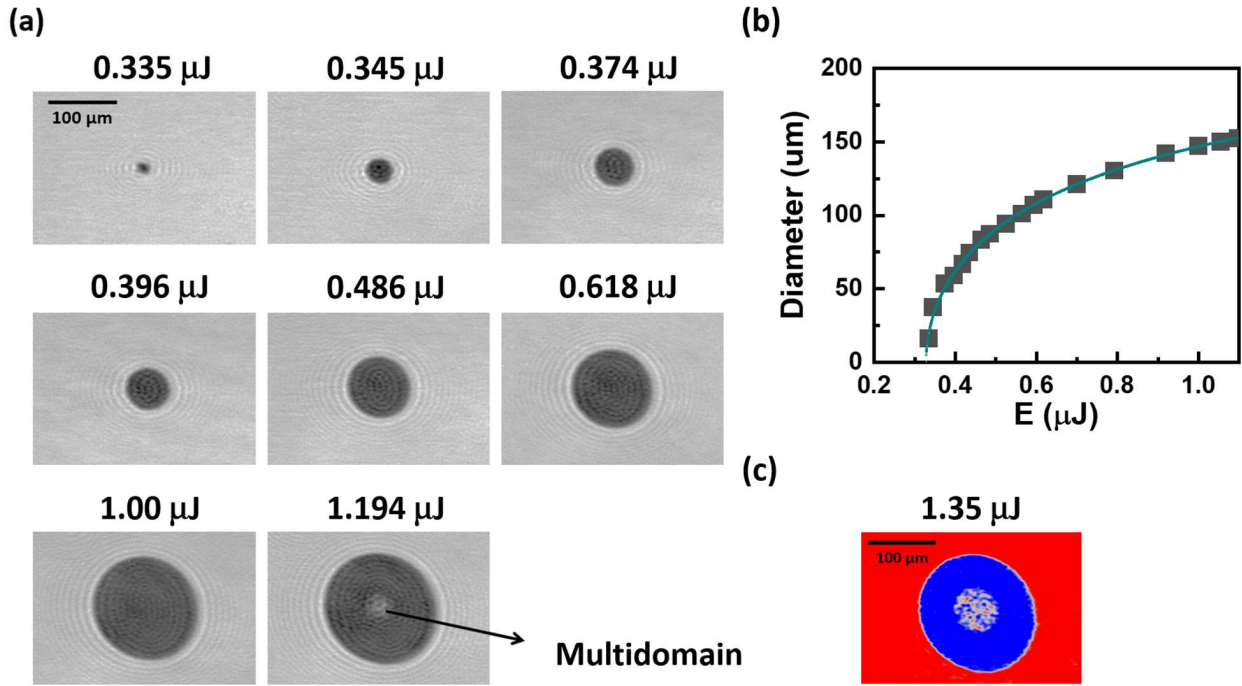

**FIG. S4.** Single pulse switching for a $Gd_{16}Co_{84}$ (5 nm) thin film. (a) Kerr images obtained after a single laser pump pulse irradiation with various pulse energy. (b) Reversed domain size as a function of the pump pulse energy, deduced from images in (a). (c) Longitudinal MOKE images were extracted at high laser energy to demonstrate the multidomain state resulting from excessive heating of the sample.

## S5. Endurance test on in-plane magnetized Gd$_{18.4}$Co$_{81.6}$ and Gd$_{21}$Co$_{79}$ alloys at room temperature

Fig. S5 shows the toggle magnetization switching after a thousand laser pump pulses.

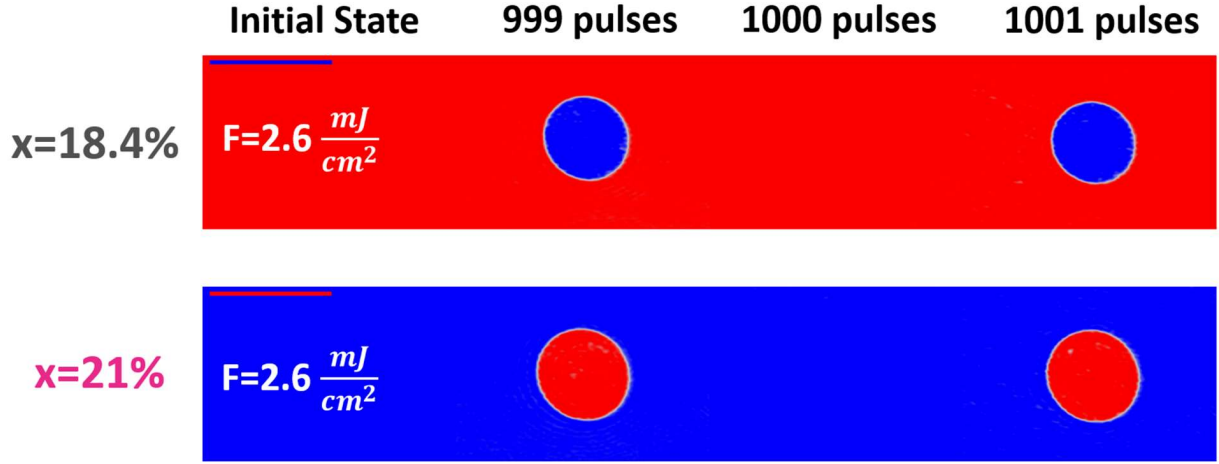

**FIG. S5.** Static longitudinal Kerr images after 999, 1000, and 1001 laser pump pulses shined on a Glass/Ta (3 nm)/Gd$_x$Co$_{100-x}$ (5 nm)/Cu (1 nm)/Pt (3 nm), with a repetition rate of 100 kHz for (a) x=18.4% and (b) x=21%. A scale bar of 100 μm is presented.

## S6. Single pulse all-optical helicity-independent switching measurements performed at the boundary between two magnetic domains

Fig. S6 shows that single pulse AO-HIS depends only on the initial magnetic configuration and the number of laser pump pulses. It also proves that the dipolar field resulting from the domain formation does not affect the magnetization reversal.

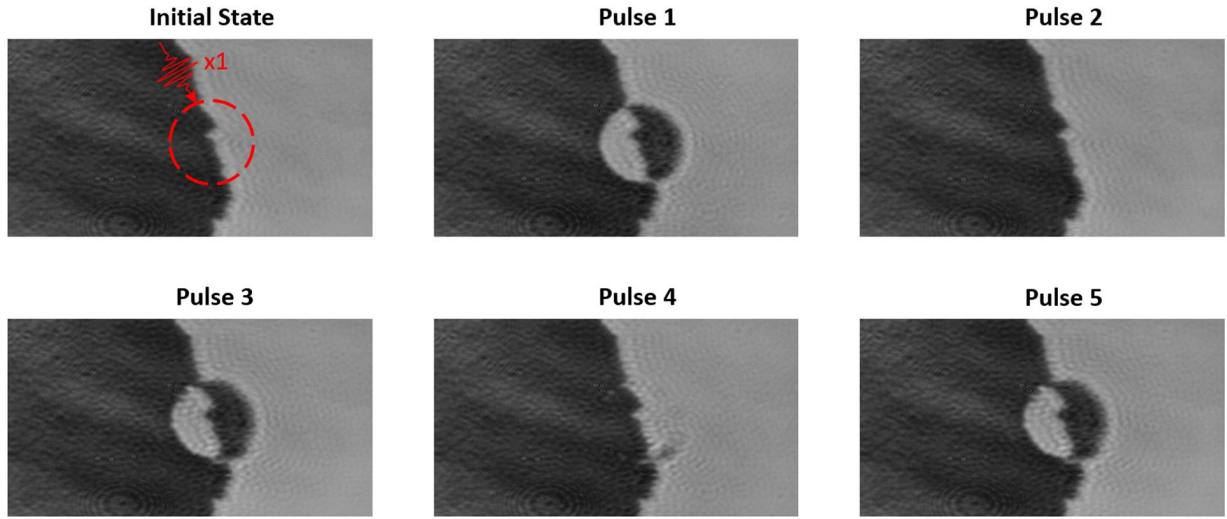

**FIG. S6.** Kerr images obtained from an initial multidomain state and after five consecutive laser pump pulses on a Glass/Ta (3 nm)/ $Gd_{15}Co_{85}$ (5 nm)/Cu (1 nm)/Pt (3 nm) sample.

## S7. Pump pulse energy-dependent optical switching measurements of $Gd_5Co_{95}$ thin film

To prove that no deterministic AO-HIS could be detected on $Gd_5Co_{95}$ film (x=5%), we show in Fig. S7 optical switching measurements without and with an applied in-plane magnetic field for various laser pump fluences. The switching of a circular area could not be observed. Neither a switching pattern nor a multidomain state was observed before burning and damaging the sample, indicating that the Curie temperature of this sample is too high to be demagnetized enough to either observe AO-HIS or heat-assisted magnetic recording. Note that the strength of the field is close but lower than the coercivity used to assist the switching. In principle, the external field should assist the switching [2]. Strikingly, only a speckled domain and no circular-shaped reversal domain were observed under applying the field, stressing again the Curie temperature ($T_C$) of this alloy is too high.

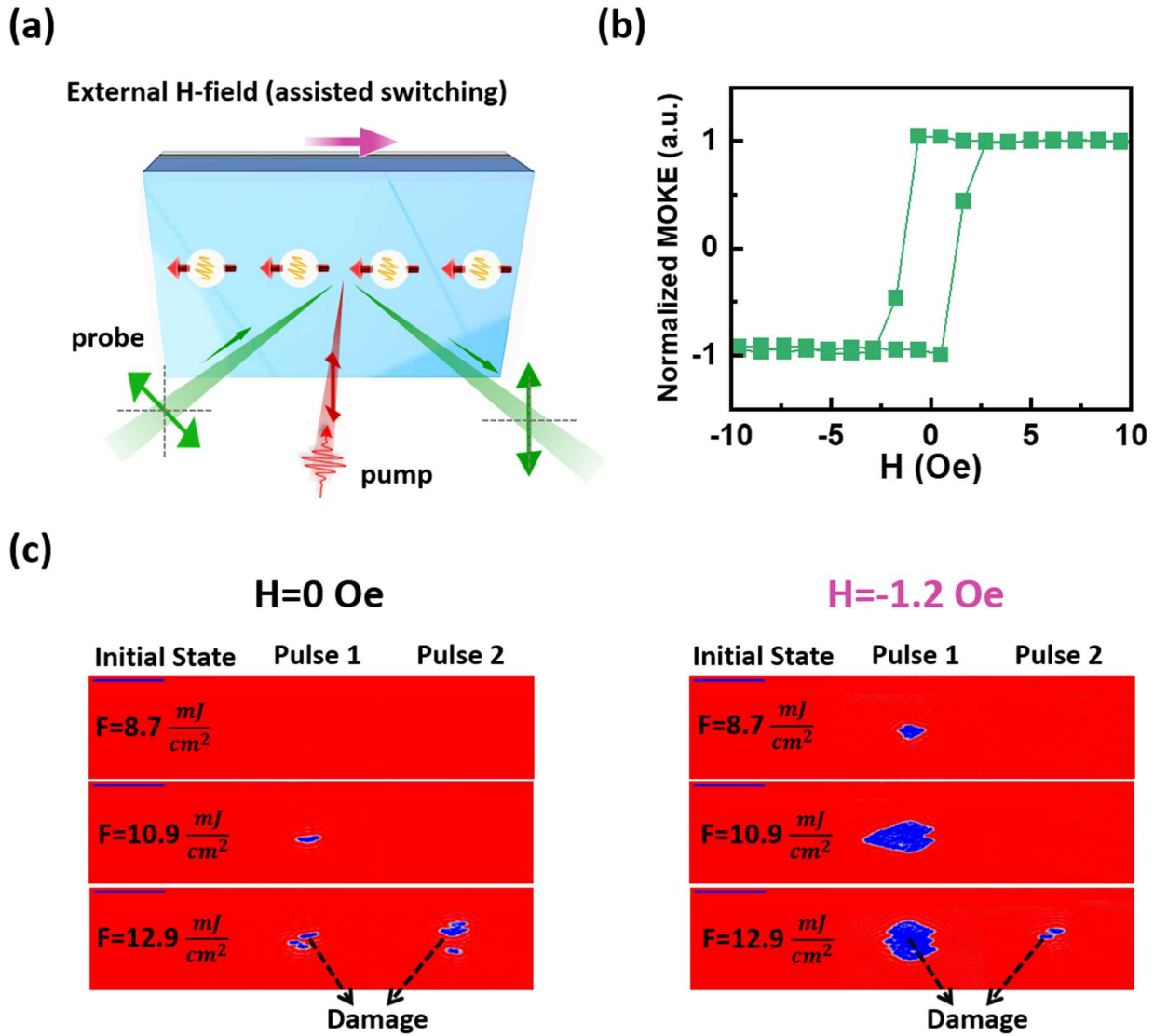

**FIG. S7.** (a) Schematic of static single pulse measurements without and with external magnetic

field along the in-plane easy axis. (b) Normalized MOKE signal as a function of the field applied along the in-plane anisotropy axis for a $Gd_5Co_{95}$ (5 nm) film. (c) Static optical switching results after 1 and 2 laser pump pulses were obtained without and with the field opposite the initial magnetization direction. A scale bar of 100 μm is presented.

## S8. Pump pulse energy-dependent optical switching measurements of $Gd_{30}Co_{70}$ thin film

We measured the optical switching without and with the external in-plane field for various laser pump fluences to demonstrate the switching behavior and the possible application of $Gd_{30}Co_{70}$ (x=30%). The reversal of a uniform domain could not be observed under a zero field; only the multidomain is formed, as shown in Fig. S8(c). However, a uniform reversed domain can be obtained by applying a field (amplitude less than the coercivity) opposite to the initial magnetization direction. The results of Fig. S8(d) show that the switching behavior depends on the amplitude of the field, which demonstrates a similar behavior as seen in heat-assisted magnetic recording [2].

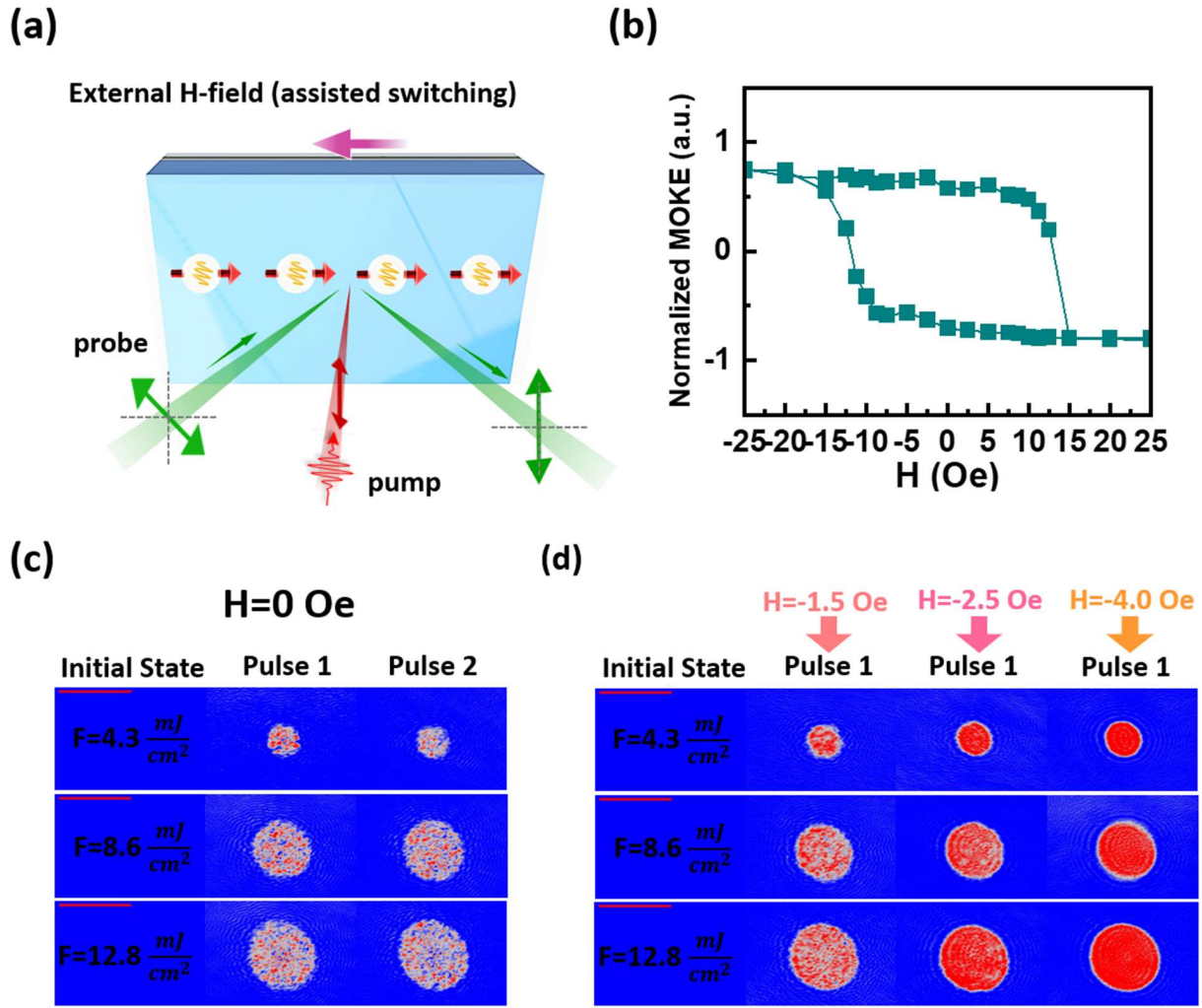

**FIG. S8.** (a) Schematic of static single pulse measurements without and with external magnetic field along the in-plane easy axis. (b) Normalized MOKE signal as a function of the field applied along the in-plane anisotropy axis for a $Gd_{30}Co_{70}$ (5 nm) film. (c) Static longitudinal Kerr images after 1 and 2 laser pump pulses under a zero field. (d) The field-assisted optical switching after a single pump pulse with different field amplitudes. A scale bar of 100 μm is presented.

## S9. Single pulse all-optical helicity-independent switching measurements performed at two orthogonal in-plane directions

Typically, in-plane magnetization could be aligned along different directions in the plane. We compared the MOKE hysteresis loops with the magnetic field applied along two orthogonal in-plane directions and measured the angle-dependent remanence. The results of Figures S9(a) and (b) confirmed the existence of well-defined in-plane anisotropy in our samples. Then, to see the AO-HIS effect on two in-plane directions, we first initialized the magnetization along the in-plane easy axis direction (corresponding to 0 degrees in Fig. S9(a)); the results show that the sample exhibits AO-HIS, as shown in Figures S9(c) and 2. Secondly, the magnetization was initialized along the in-plane hard axis direction (corresponding to 90 degrees in Fig. S9(a)), and then micro-sized domains were created under a zero field upon single laser pump pulse irradiation. The results of Figures S9(d) and (e) show that those domains can still be reversed partially in a repeatable way and are insensitive to the multidomain surrounding them. The results of Figures S9(c)-(e) clearly summarize that the switching with a circular area only appears when the magnetization is initialized along the direction of the in-plane easy axis, indicating that having the well-defined in-plane magnetic anisotropy is also a key ingredient to observe well-defined reversed domain.

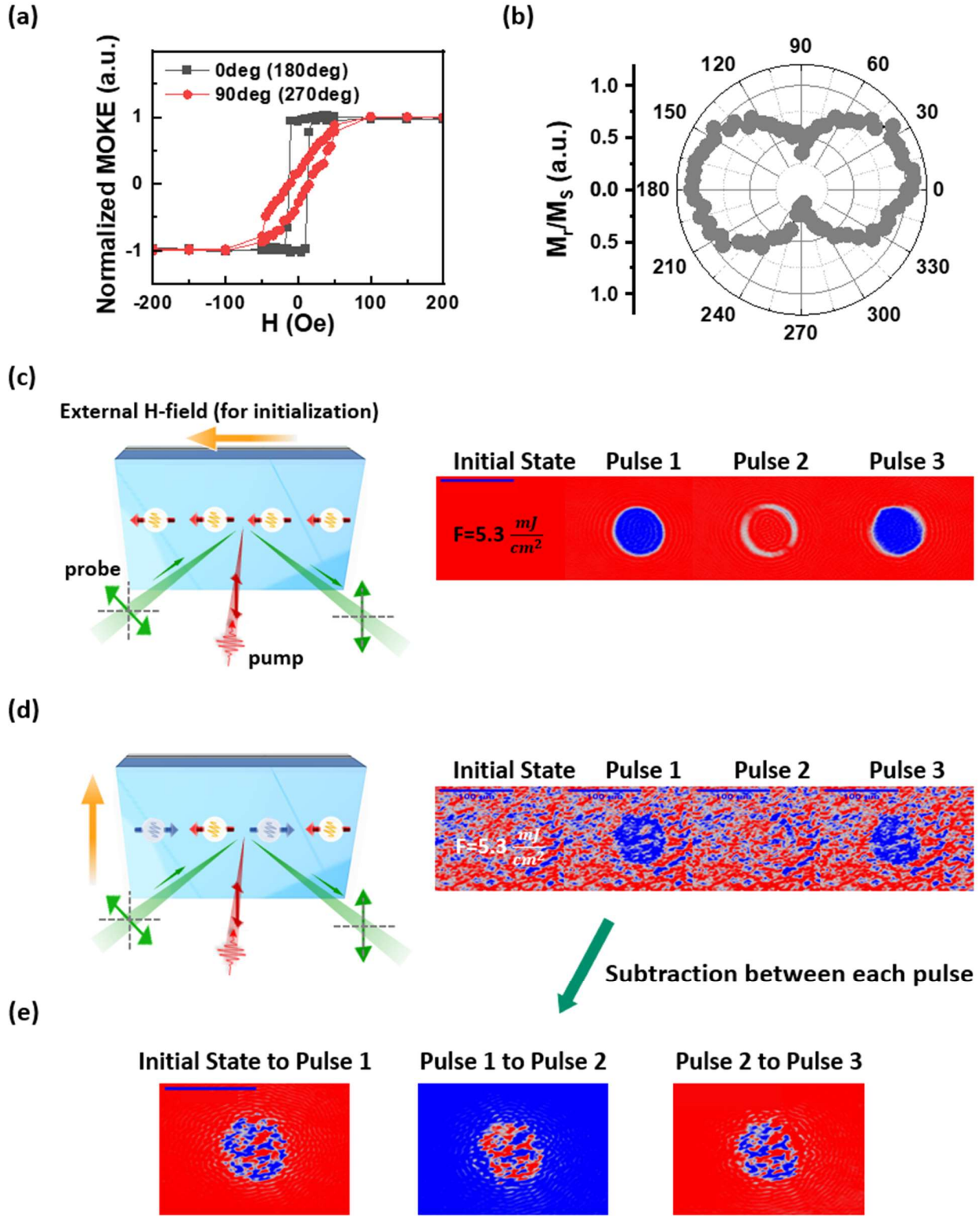

**FIG. S9.** (a) Normalized longitudinal MOKE hysteresis loops measured for the magnetic field applied along the longitudinal (0 deg) and transversal (90 deg) in-plane direction. Here the sample of Glass/Ta (3 nm)/ $Gd_{15}Co_{85}$ (5 nm)/Cu (1 nm)/Pt (3 nm) was used for demonstration. (b) Angle-

dependent remanent ratio $M_r/M_S$ of a $Gd_{15}Co_{85}$ film which was collected by a vibrating-sample magnetometer. Static longitudinal Kerr images after 1, 2, and 3 pump pulses when the field initializes the magnetization along the (c) easy axis and (d) hard axis in the plane. A scale bar of 100 μm is presented. (e) From left to right are the pictures taken from (d) showing the difference between each pump pulse, as indicated by the label above the MOKE images.

## S10. Comparison of the demagnetization threshold and Curie temperature

Previous theoretical works have suggested that a multidomain state will be formed during the successive slow cooldown when the lattice temperature rises above the $T_C$ of GdCo alloys [3]. This implies that the multidomain relates to the $T_C$ of a material. Here, we define the pump pulse energy required to observe multidomain patterns in the irradiation area as a demagnetization energy threshold ($E^{th}_{Dem}$). Accordingly, $E^{th}_{Dem}$ versus Gd concentration is compared to the $T_C$ of each Gd concentration extracted from the literature [4]. Both cases follow a similar trend.

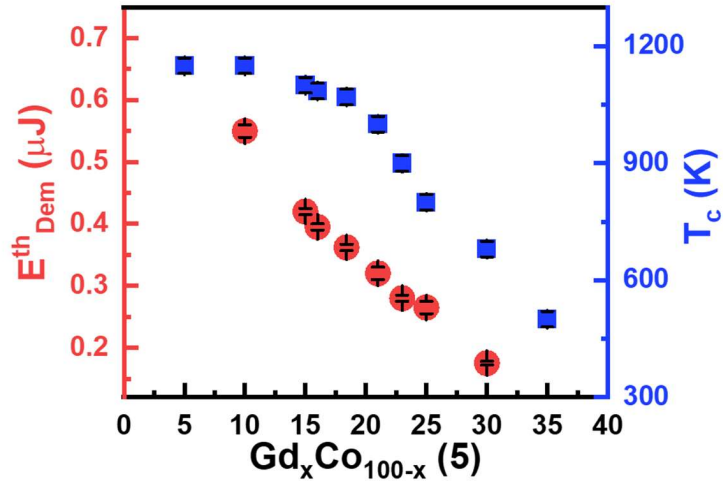

**FIG. S10.** The demagnetization threshold ($E^{th}_{Dem}$) energies compared to literature values of $T_C$ as a function of Gd concentration.

## S11. MOKE hysteresis loop measurements of in-plane magnetized Gd$_{25}$Co$_{75}$ with different thicknesses

The longitudinal MOKE hysteresis loops were obtained by applying a magnetic field along the in-plane easy axis. In all cases, square hysteresis loops with ~100% remanence for H reveal the existence of well-defined in-plane magnetic anisotropy in all Gd$_{25}$Co$_{75}$ films. The main hard axis also sits along the direction normal to the sample surface.

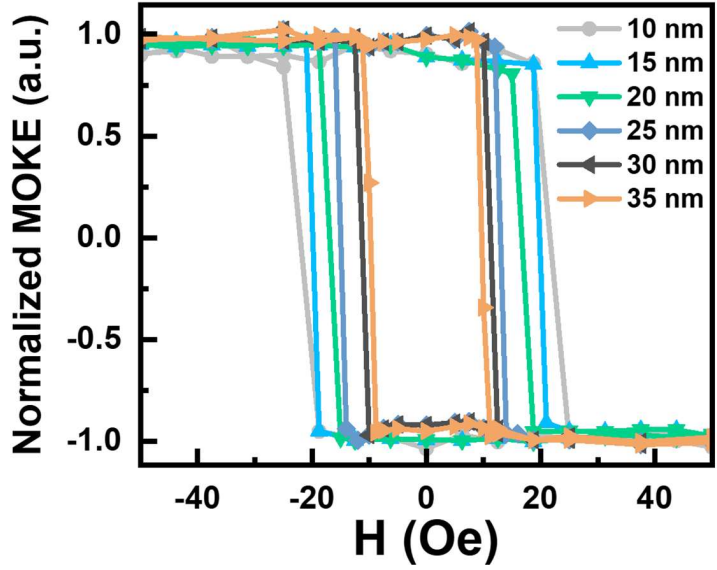

**FIG. S11.** Normalized longitudinal MOKE hysteresis loops of Glass/Ta (3 nm)/Gd$_{25}$Co$_{75}$ (t nm)/Cu (1 nm)/Pt (3 nm) for various thicknesses (t) ranging from 10 nm to 35 nm with a field applied in-plane along the easy axis.

## S12. The calculation of the absorption profile of Gd$_{25}$Co$_{75}$ with various thicknesses

The multilayer structure of Glass/Ta (3 nm)/Gd$_{25}$Co$_{75}$ (t nm)/Cu (1 nm)/Pt (3 nm) was used to study the thickness-dependent single pulse AO-HIS. In order to understand the amount of energy absorbed by the sample, it is interesting to know the optical absorption profiles as a function of the sample depths [5-7]. As shown in Fig. S12(a), the absorption profile in GdCo is almost constant for t<20 nm, which indicates that the entire GdCo layer can be uniformly demagnetized. By contrast, when t≥20 nm, the absorption profile starts to become non-uniform. Fig. S12(b) shows the total amount of laser energy deposited in the GdCo layer. The results reveal that the absorption reaches saturation at t~20 nm, and then almost stays constant. The result means that when the thickness is greater than 20 nm, the total energy injected into the GdCo is nearly maintained, giving rise to a temperature gradient in the magnetic layer. This indicates that the front part of the sample absorbs most of the laser energy, resulting in an insufficient demagnetization in the deeper part of the GdCo layer. Accordingly, we attribute a maze domain structure for t=35 nm to the significant thermal gradient in the GdCo layer.

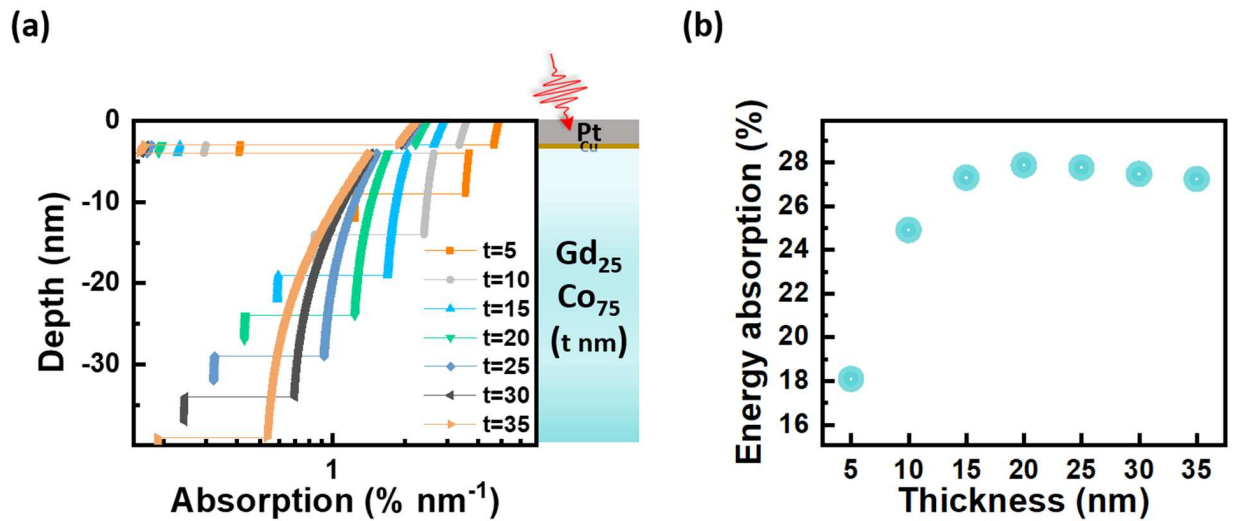

**FIG. S12.** (a) Calculated optical absorption as a function of depth for different Gd$_{25}$Co$_{75}$ film thicknesses. (b) Total energy absorption in the Gd$_{25}$Co$_{75}$ layer as a function of film thickness. The optical indices used were obtained from a previous study [7].

## S13. The definition of the time needed for T1 and T2

From the time-resolved magnetization dynamic curve, three typical features can be determined by their slopes, as observed by previous works [8,9]. The magnetization dynamics of the $Gd_{18.4}Co_{81.6}$ alloy is taken as an example to demonstrate the way to extract T1 and T2, as shown in Fig. S13(a); three features can be described as follows: (1) a fast initial drop as a result of ultrafast demagnetization (black curve; the slope is the highest among the three). (2) a recovery in the opposite direction of the initial state (orange curve; the slope is the intermediate among three). (3) a plateau where the value of normalized MOKE stays almost constant (dark cyan curve; the slope is nearly zero). By obtaining those three features (slopes), we can define the time scale "T1" at which the transition between features (1) and (2) happens. Similarly, "T2" can be determined by the time scale at which the transition between features (2) and (3) happens. By carefully analyzing the time-resolved MOKE traces for various Gd concentrations (Fig. S13(b)), we can summarize T1 and T2 as a function of Gd concentration, as shown in Fig. 4(c).

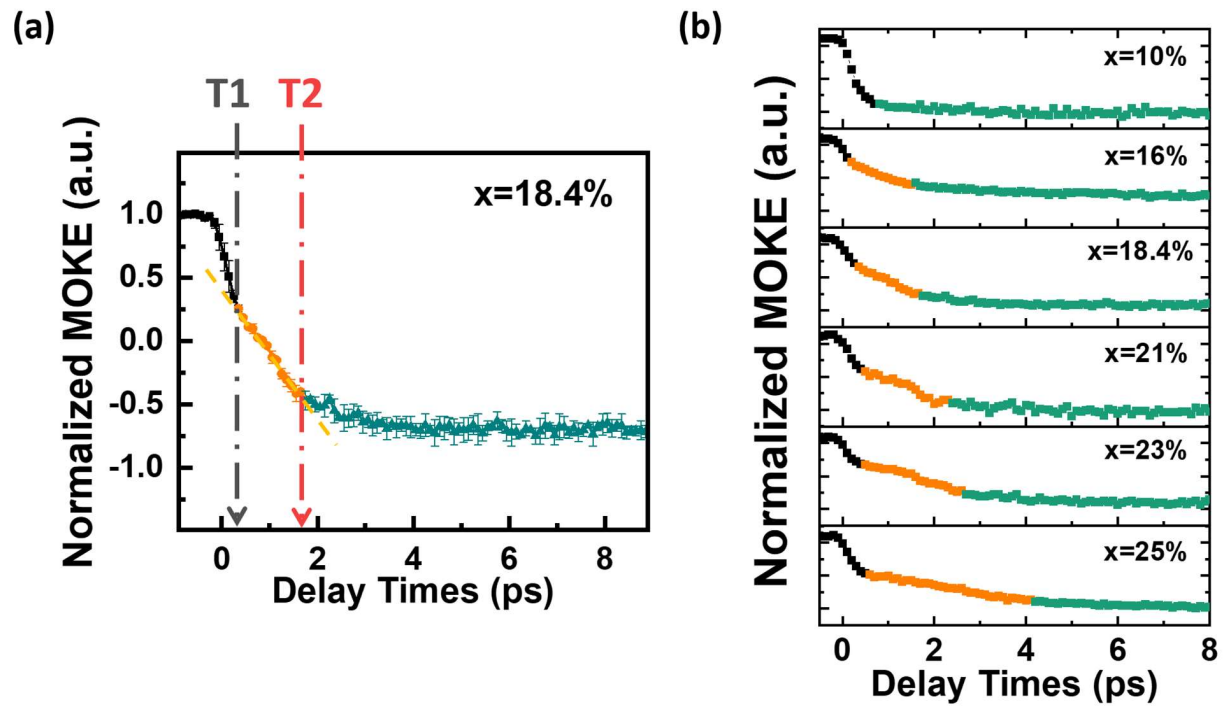

**FIG. S13.** (a) Time-resolved MOKE trace of $Gd_{18.4}Co_{81.6}$ alloy is taken as an example to define the time at which for T1 and T2. (b) Three different features determined by their slopes are plotted as a function of Gd concentration.